**Academic Ranking with Web Mining and Axiomatic Analysis**

[1]Kun Tang, [2]Qiwei Jin, [2]Xin Zou*, [1]Jiansheng Yang, [3]Michael Vannier, [4]Ge Wang*

[1]School of Mathematical Sciences
Peking University
5 Yi He Yuan Street, Beijing 100871, China
kun-tang@pku.edu.cn
jsyang.rena@gmail.com

[2]Microsoft Academic Search
5 Danling Street, Haidian District
Beijing, 100085, China
qiwjin@microsoft.com
xinz@microsoft.com

[3]Department of Radiology
University of Chicago Medical Center
5841 South Maryland, Q-226, MC2026
Chicago, IL 60637, USA
mwvannier@gmail.com

[4]School of Biomedical Engineering and Sciences
Virginia Tech
Blacksburg, Virginia 24061, USA
ge-wang@ieee.org

*Corresponding authors: Xin Zou and Ge Wang

December 24, 2012

**Abstract**

Academic ranking is a public topic, such as for universities, colleges, or departments, which has significant educational, administrative and social effects. Popular ranking systems include the US News & World Report (USNWR), the Academic Ranking of World Universities (ARWU), and others. The most popular observables for such ranking are academic publications and their citations. However, a rigorous, quantitative and thorough methodology has been missing for this purpose. With modern web technology and axiomatic bibliometric analysis, here we perform a feasibility study on Microsoft Academic Search metadata and obtain the first-of-its-kind ranking results for American departments of computer science. This approach can be extended for fully automatic intuitional and college ranking based on comprehensive data on Internet.

**Introduction**

Whether it is good or bad, gaining a better rank in the US News Report & World Report (USNWR) is a priority of many university administrators and faculty members. This rank is widely used, such as for students to select universities. Other ranking systems and reports are also of reference value; for example, the Academic Ranking of World Universities (ARWU).

The current ranking methods are based on survey, analysis and synthesis. Hence, subjective opinions play a major role. Various weighting criteria lead to different ranking lists. One can select those lists on which preferred outcomes are seen, and even make lobbying efforts to induce favorable scoring. Clearly, the inconsistency and confusion in practice compromise the credibility and impact of the current academic ranking results. Here we suggest a new ranking methodology that utilizes comprehensive web resources, credit team members/co-authors axiomatically, and quantify academic outputs objectively and rationally. As an initial effort to demonstrate the feasibility and utility, we focus on ranking American departments of computer science.

In the field of computer science, the number of annual publications (including journal and conference papers) has been greatly increased from 10,000 forty years ago to over 200,000 now. The average number of coauthors has been increased from 1.25 to 3.12 over the past 50 years. Some papers may even have more than 100 co-authors. Also, it is known that authors tend to cite their own work more often. Hence, how to assign credits to co-authors and how to exclude self-citation must be solved for least biased academic assessment.

In 2005, Hirsch defined the *h*-index as a bibliometric indicator (Hirsch 2005). Inspired by this work, various new bibliometric indicators were developed (Hagen 2008; Zhang 2009). Most of these indicators do not differentiate coauthors' relative contributions. There are two popular approaches for crediting coauthors. The first one lets each co-author receive the full credit. The second one gives every coauthor an equal credit. These measures are evidently too rough, since co-authors' contributions to a paper can be rather uneven. The harmonic allocation method (Hagen 2008) was designed to overcome the unfairness. In this scheme, the weight of the *k*-th co-author is subjectively set to $\frac{1}{k}/\sum_{i=1}^{n}\frac{1}{i}$, where *n* is the number of co-authors. An alternative credit sharing method (Zhang 2009) was proposed based on some arguable heuristics. Hirsch suggested a *hbar* index to take into account the effect of multi-authorship (Hirsch 2010). Nevertheless, the *hbar*-index does not extract coauthors' credit shares on any specific paper. On the other hand, the axiomatic credit-sharing scheme (Wang and Yang 2010) is a novel solution, which is referred to as the *a*-index since it was axiomatically derived.

Here we refine the number of citations with the *a*-index and exclude self-citations proportionally. After a co-author of a paper receives an appropriate credit according to the *a*-index, he or she will obtain his or her own share of the total number of citations to that paper. Citations will be excluded from one's share of a paper to his/her share of another paper. Then, we define the *ah*-index such that a co-author has an *ah*-index value *x* if he or she has at most *x* papers to which his or her pure share of the total number of citations is at least *x*. With these sophisticated refinements and huge amounts of web-based metadata, the research output can be convincingly quantified as a foundation for fair and open ranking.

**Prior Art**

Since 1983, the US News & World Report (USNWR) keeps publishing annual listing of American Best Colleges. Inspired by USNWR, other ranking results emerged using different methods. There are now more than 50 different systems for ranking institutions. Most of these rankings use the weighted-sum mechanism. They rely on some relevant (correlated, to different degrees) indicators, and use the sum of weighted scores to determine the rank of an institution.

The Academic Ranking of World Universities (ARWU) is a good example, with annual ARWU data available since 2003. In the field of computer science, the ranking relies on the five bibliometric indicators (http://www.shanghairanking.com/ARWU-SUBJECT-Methodology-2011.html): (1) Alumni (10%), as quantified by the number of alumni winning Turning Awardees since 1961; (2) Award (15%), the number of faculty winning Turning Awardees since 1961; (3) HiCi (25%), the number of highly cited papers; (4) PUP (25%), the number of papers indexed in the Science Citation Index (SCI); and (5) TOP (25%), the percentage of papers published in the top 20% journals in the field. In each category, the university with the maximum score receives 100 points, and the other universities are measured in terms of percentages relative to the maximum score. The total credit for a university is a weighted sum of the five measures.

In addition to the above indicators, there are other variants and features (Eom 1994; Zhou 2005; Jason, Pokorny et al. 2007; Docampo 2012). It is highly non-trivial how to select from indicators and how to weight them. With rapid development of web science and technology, it would be ideal to have an intuitional ranking system applying contemporary web-based data-mining techniques to ever-expanding digital contents for authoritative ranking results.

## Methods

Scientific publication is a main outcome of research and development, and the number of citations is a well-accepted key observable on the impact of a paper. Among all the ranking systems, publications and citations have been extremely important indicators but the scientific credit of a paper has not been individually assigned and comprehensively analyzed in the context of academic ranking. Here we make the first attempt to accomplish this goal.

Let us first describe how to calculate an individual co-author's credit in a specific paper using our axiomatic approach (Wang and Yang 2010). The axiomatic system consists of the three axioms: (1) <u>Ranking Preference</u>: a better ranked co-author has a higher credit; (2) <u>Credit Normalization</u>: the sum of individual credits equals 1; and (3) <u>Maximum Entropy</u>: co-authors' credit shares are uniformly distributed in the space defined by Axioms 1 and 2. As a result, if there is no evidence that some co-authors made an equal contribution, then the $k$-th co-author of a paper by $n$ co-authors has a credit share $\frac{1}{n}\sum_{j=k}^{n}\frac{1}{j}$. If the last author is the corresponding author, he or she can be considered as important as the first author. If there are no other two co-authors who have the same amount of credit, then the first or last authors credit's is $\frac{1}{n-1}\sum_{j=1}^{n-1}\frac{1}{j+1}$, and the $k$-th co-author's credit is $\frac{1}{n-1}\sum_{j=k}^{n-1}\frac{1}{j+1}$, $k \neq 1$ and $k \neq n$. For the algorithmic details (Wang and Yang 2010), see Figure 1.

**Algorithm 1: Pseudo Code for Computing a Co-author's Credit**

**function** Credit (author, paper) **return** real
  k <- author's position in the list of co-authors
  n <- number of co-authors
  **if** the last author is the corresponding author **then**
    **if** k = 1 or k=n **then**
    **return** $\frac{1}{n-1}\sum_{j=k}^{n-1}\frac{1}{j+1}$.
    **else**
    **return** $\frac{1}{n}\sum_{j=k}^{n}\frac{1}{j}$

Figure 1. Pseudo code for computing a co-author's credit in a paper.

Aided by the *a*-index, the number of citations to a paper can be proportionally assigned to each co-author. In other words, a co-author with an *a*-index value *c* for a paper being cited *M* times gains *c*M* citations to that paper, which is referred to as the *ac*-index. To be more objective, it is preferred that self-citations be removed from the citations to a paper. When a researcher published a paper, his or her institution gets a credit. The credit for an institution can be measured as the sum of the credits earned by those co-authors who are with the institution. Aided by the *a*-index, we can exclude self-citations specific to individual co-authors and their related contributions, as shown in Figure 2. Then, we can define the institutionally-oriented *ah*-index using the algorithm in Figure 3.

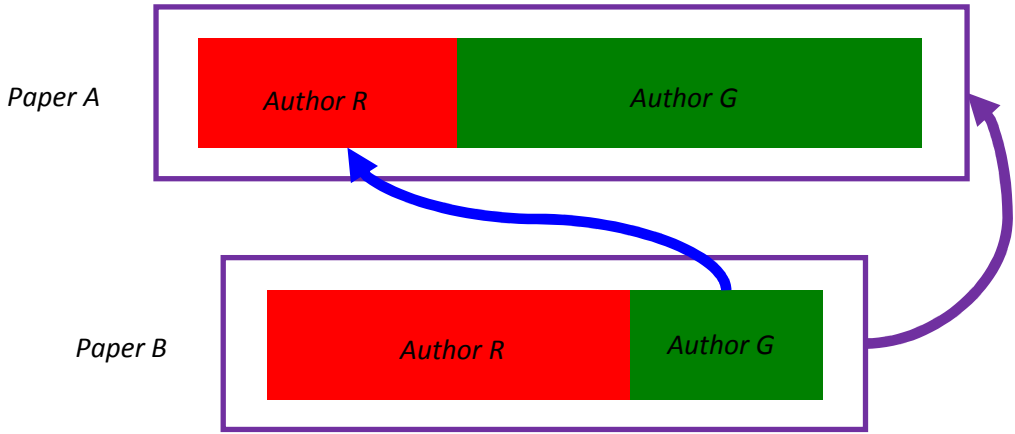

Figure 2. Axiomatic exclusion of self-citation. The citation (indicated by the purple arrow) to Paper A from Paper B is a self-citation because an author R has his/her red shares in both the papers. The pure self-citation can be excluded if we use the axiomatic strength of the citation to the author R's share in Paper A from the other author G's share in Paper B (indicated by the blue arrow).

Figure 3. Pseudo code for computing an institutional credit from all the involved papers.

## Data Source

Microsoft Research performs basic and applied research in computer science and software engineering in more than 50 areas. It has expanded to eight locations worldwide with collaborative projects. Microsoft Academic Search (MAS) (http://academic.research.microsoft.com) is a free service of Microsoft Research to help study academic content. This service not only indexes academic papers but also reveals relationships among subjects. Under this service, the number of publications is more than 40 millions, and the number of authors more than 18.9 millions. Thousands of new papers are integrated into the database regularly. In the domain of computer science, there are more than 6-million papers. About 40% of them are from journals. About 35% of them are from conference proceedings. The other papers do not have a clear association with either a journal or a conference.

In MAS, search results are sorted, covering the entire spectrum of science, technology, medicine, social sciences, and humanities. The current partners are dozens of publishers and other content providers. The novel analytic features include the **genealogy graph** for advisor-advisee relationships based on the information mined from the web and user input, the **paper citation graph** showing the citation relationships among papers, the organization comparison in different domains, **author/organization rank lists**, the **academic map** presenting organizations geographically, the **keyword detail** with the Stemming Variations and Definition Context.

The MAS software was mainly written in C/C++/C#/SQL/ASP.net on a dedicated system consisting of the following modules: Offline Data Processing, Metadata Extraction, Reference Building, Name Disambiguation, Online Index Building/Servicing, Data Presentation, and tools to support users' feedback and contribution. MAS has a heterogeneous computer network with Dell computer workstations for routine use (dual-core 2.53GHz CPU, 64G RAM).It took 5 days for us to process 1-million papers from collection of them to completion of the whole procedure including metadata extraction, citation context extraction, reference matching within the 1-million papers and citation analysis between the existing papers and newly added papers. This system is capable of handling up to 100-million documents using existing hardware and software.

In this project, we collected all the information from MAS metadata, computed individual credits and excluded self-citations using the above-described algorithms. Currently, MAS does not collect non-English publications, which will be very likely included in the future. An author's information included his/her institution. The email address was extracted from the PDF file or other electronic publications. Such information was cross-checked with data mined from his/her academic homepage. Also, a user can make corrections or provide metadata using built-in tools. An automatic module was developed to analyze coauthors' names to eliminate any ambiguity in the cases of the same person with multiple email addresses, for different working organizations, by various name spellings, different individuals with the same name, and so on. When there was any error in the metadata, the whole entry was removed; for example, if some author information was not extracted successfully, the publication would be discarded. In other words, we only used the successfully preprocessed information. A computer science paper receives a credit from a citing paper which is not necessarily also a computer science paper. The corresponding author is not easy to identify in the current database. In the calculation, if one co-author provided his/her email in the paper, we treated him/her as the corresponding author. It is not the best solution but it is fair for all the institutions and already far more comprehensive and rigorous than the current ranking systems.

**Results**

We calculated the ranks of American departments of computer science by our $ac$- and $ah$-indices, the $aj$-index that is defined as the sum of the $a$-index weighted by the journal impact factor for each of all the papers associated with a department, and the $aac$-index defined as the averaged $ac$-index. Table 1 shows the relevant ranks by each of all these measures. The $ac$-index-based ranking reflects the overall impact in terms of "pure" citations from a department, and emphasized in Table 1. The $acc$-index-based ranking is after the normalization with respect to the number of coauthors associated with a department. The $ah$-index-based results represent a refinement to the $h$-index-based ranking. The $aj$-index is advantageous in terms of promptness; that is, no need to wait for citations.

The Spearman and Kendall correlation data are in Tables 2 and 3 for the data from top 50 American universities ranked by USNWR. The reason for the use of Kendall and Spearman correlation measures, instead of the Pearson correlation coefficient, is to capture the correlative relationships better among trends in terms of different bibliometric indicators, since these relationships are not always linear; for example, the $ac$-index is proportional to the square of the $ah$-index.

| ac-Rank | Institution | ac-index | aac-index | ah-index | aj-rank (2012)[1] | # of authors | # of papers | ARWU[2] (2011) | USNWR[3] (2010) |
|---|---|---|---|---|---|---|---|---|---|
| 1 | Massachusetts Institute of Technology | 274440.5 | 48.1 | 197 | 1 | 5711 | 43701 | 2 | 1 |
| 2 | Stanford University | 267123.6 | 50.7 | 205 | 2 | 5266 | 45798 | 1 | 1 |
| 3 | Carnegie Mellon University | 234860.7 | 56.8 | 170 | 9 | 4137 | 42258 | 6 | 1 |
| 4 | University of California Berkeley | 234236.7 | 53.3 | 194 | 3 | 4397 | 39679 | 3 | 1 |
| 5 | University of Illinois Urbana Champaign | 130772.0 | 34.7 | 129 | 4 | 3765 | 33008 | 11 | 5 |
| 6 | Georgia Institute of Technology | 102320.4 | 27.5 | 112 | 11 | 3719 | 30509 | 19 | 10 |
| 7 | University of Maryland | 90477.97 | 33.0 | 117 | 12 | 2740 | 25523 | 12 | 14 |
| 8 | University of California Los Angeles | 81258.45 | 29.2 | 113 | 6 | 2786 | 24257 | 17 | 14 |
| 9 | University of Michigan | 77306.04 | 23.1 | 104 | 8 | 3343 | 23993 | 14 | 13 |
| 10 | University of Southern California | 76389.19 | 27.7 | 102 | 14 | 2759 | 25760 | 9 | 20 |
| 11 | University of Washington | 75294.52 | 25.0 | 116 | 13 | 3016 | 22242 | 16 | 7 |
| 12 | University of Texas Austin | 73734.15 | 22.9 | 107 | 15 | 3224 | 26996 | 8 | 8 |
| 13 | Cornell University | 72117.64 | 36.2 | 117 | 28 | 1994 | 16518 | 7 | 5 |
| 14 | University of Wisconsin Madison | 65272.32 | 28.6 | 113 | 21 | 2281 | 16485 | 41 | 11 |
| 15 | University of California San Diego | 64355.73 | 21.9 | 102 | 5 | 2934 | 25860 | 13 | 14 |
| 16 | University of Minnesota | 59021.07 | 22.7 | 92 | 10 | 2604 | 18725 | 34 | 35 |
| 17 | Columbia University | 57890.46 | 30.9 | 91 | 16 | 1873 | 16475 | 17 | 17 |
| 18 | Princeton University | 57189.62 | 44.8 | 104 | 20 | 1276 | 14645 | 4 | 8 |
| 19 | Purdue University | 56405.08 | 20.0 | 92 | 19 | 2814 | 22403 | 15 | 20 |
| 20 | University of Massachusetts Amherst | 54316.84 | 28.8 | 103 | 45 | 1889 | 15288 | 30 | 20 |
| 21 | University of California Irvine | 51333.04 | 28.7 | 89 | 24 | 1790 | 16958 | 21 | 28 |
| 22 | University of Pennsylvania | 50660.41 | 31.3 | 90 | 17 | 1616 | 13004 | 28 | 17 |
| 23 | Rutgers University | 49438.86 | 31.0 | 92 | 25 | 1595 | 15981 | 25 | 28 |
| 24 | California Institute of Technology | 45189.05 | 33.4 | 88 | 23 | 1352 | 9658 | 10 | 11 |
| 25 | Harvard University | 42441.6 | 16.5 | 83 | 7 | 2571 | 14138 | 5 | 17 |
| 26 | Pennsylvania State University | 38848.23 | 15.2 | 71 | 26 | 2564 | 18193 | 41 | 28 |
| 27 | University of California Santa Barbara | 36009.39 | 25.3 | 74 | 37 | 1425 | 11964 | 27 | 35 |
| 28 | University of North Carolina Chapel Hill | 35917.43 | 31.4 | 80 | 39 | 1144 | 8830 | 22 | 20 |
| 29 | Ohio State University | 34019.76 | 16.1 | 67 | 33 | 2110 | 15015 | 28 | 28 |
| 30 | University of Colorado Boulder | 33237.4 | 22.4 | 74 | 41 | 1485 | 10236 | 26 | 39 |
| 31 | Yale University | 28887.68 | 27.4 | 69 | 18 | 1056 | 8760 | 20 | 20 |
| 32 | Texas A&M University | 28474.24 | 13.3 | 57 | 27 | 2141 | 14216 | 41 | 47 |
| 33 | Rice University | 26423.15 | 32.6 | 75 | 47 | 811 | 7948 | 34 | 20 |
| 34 | New York University | 26142.26 | 25.0 | 73 | 42 | 1045 | 8531 | 34 | 28 |
| 35 | University of Virginia | 26021.63 | 20.8 | 64 | 48 | 1252 | 8426 | 32 | 28 |
| 36 | University of California Davis | 25739.79 | 15.6 | 69 | 29 | 1647 | 11589 | 30 | 39 |
| 37 | Brown University | 25208.12 | 32.7 | 70 | 46 | 771 | 7975 | 34 | 20 |
| 38 | Northwestern University | 25198.38 | 18.6 | 60 | 35 | 1353 | 11347 | 33 | 35 |
| 39 | Duke University | 24907.47 | 17.9 | 62 | 31 | 1389 | 10625 | 24 | 27 |
| 40 | Johns Hopkins University | 24738.61 | 15.6 | 63 | 34 | 1582 | 10999 | NR[4] | 28 |
| 41 | Boston University | 24193.62 | 22.1 | 68 | 32 | 1097 | 9774 | 41 | 47 |
| 42 | Washington University in St. Louis | 22161.58 | 21.0 | 65 | 30 | 1057 | 7645 | NR[4] | 39 |
| **43** | Rensselaer Polytechnic Institute | 21734.5 | 17.0 | 60 | 44 | 1280 | 9449 | NR [4] | 47 |
| **44** | Virginia Tech | 20701.25 | 9.5 | 53 | 36 | 2180 | 13664 | NR [4] | 44 |
| **45** | University of Arizona | 20694.63 | 12.7 | 58 | 38 | 1632 | 10419 | 34 | 47 |
| **46** | Stony Brook University | 20471.27 | 26.6 | 56 | 49 | 770 | 7400 | NR [4] | 44 |
| **47** | University of Florida | 20040.97 | 10.2 | 50 | 22 | 1960 | 13455 | 34 | 39 |
| **48** | University of Rochester | 19451.28 | 25.7 | 67 | 50 | 756 | 5965 | NR [4] | 47 |
| **49** | University of Utah | 17729.43 | 14.7 | 56 | 40 | 1205 | 7915 | 34 | 39 |
| **50** | Dartmouth College | 14487.19 | 26.7 | 50 | 51 | 543 | 4211 | NR [4] | 44 |
| **51** | University of Chicago | 13922.64 | 18.4 | 55 | 43 | 758 | 5644 | 41 | 35 |
| **52** | University of North Carolina Charlotte | 9049.804 | 17.2 | 29 | 52 | 525 | 3561 | 23 | 47 |

**ac-index**:   total of citation shares accumulated for a department after exclusion of self-citation;
**aac-index**:   averaged ac-index with respect to the number of coauthors associated with a department;
**ah-index**:   h-index of each department after associated citation counts are weighted by corresponding a-indices;
**aj-index**:   total of a-index-based credit shares weighted by journal impact factors for a department.

[1] Data collected after 1975 when the journal impact factor was introduced;
[2] http://www.shanghairanking.com/SubjectCS2011.html;
[3] http://grad-schools.usnews.rankingsandreviews.com/best-graduate-schools/top-science-schools;
[4] Not ranked.

Table 2. Spearman correlation among competing ranks.

| Spearman correlation | ac-index | aac-index | ah-index | aj-index | USNWR | ARWU |
|---|---|---|---|---|---|---|
| ac-index | 1 | | | | | |
| aac-index | 0.6478 | 1 | | | | |
| ah-index | 0.9622 | 0.7383 | 1 | | | |
| aj-index | 0.8349 | 0.3606 | 0.7572 | 1 | | |
| USNWR | 0.8704 | 0.7082 | 0.8835 | 0.7284 | 1 | |
| ARWU | 0.7858 | 0.5662 | 0.7635 | 0.7185 | 0.8080 | 1 |

Table 3. Kendall correlation among competing ranks.

| Kendall correlation | ac-index | aac-index | ah-index | aj-index | USNWR | ARWU |
|---|---|---|---|---|---|---|
| ac-index | 1 | | | | | |
| aac-index | 0.4570 | 1 | | | | |
| ah-index | 0.8496 | 0.5495 | 1 | | | |
| aj-index | 0.6696 | 0.2232 | 0.5782 | 1 | | |
| USNWR | 0.7056 | 0.5431 | 0.7271 | 0.5493 | 1 | |
| ARWU | 0.5985 | 0.4136 | 0.6022 | 0.5368 | 0.6600 | 1 |

**Discussions and Conclusion**

It can be seen in Tables 1-3 that the compared ranking systems are quite different, with the range [0.3606, 0.9622] for Spearman correlation and the range [0.2232, 0.8496] for Kendall correlation. Given the dominating status and objective nature of the scientific publications and associated others' citations among all the observable variables for institutional assessment, we believe that the *ac*-index is a most important value for institutional ranking, and the *aac*-index can be easily derived after the normalization with respect to the size of an involved team of coauthors. The *ah*-index is a somehow convenient but quite approximate proxy. The *aj*-index is an indirect measure, since the journal impact factor cannot precisely predict the impact of a particular paper. ***It is very interesting that our data consistently show that the USNWR system is clearly better than the ARWU system.***

There are large changes in rankings around a middle range among *ac*-index, *aj*-index and USNWR. Some of the changes were not unexpected. What happened that could explain such upheavals? Responsible factors might include historical reputation, total funding, student selectivity and number, and other factors used in traditional ratings. While the USNWR ranking relies on proprietary data, the AWRU ranking is more objective. In contrast to both of these rankings, our approach offers a much wider coverage of relevant data, allows a significantly higher level of mathematical sophistication, and promises a new ranking system for assessment of academic units, such as universities, institutes, colleges, departments, and research groups.

Currently, our ranking system analyzes publications only, and carries the weaknesses that cannot be addressed by publications. Complementary features are needed for improvement; for example, profits generated by spin-off companies, royalties from licensing, and other monetary amounts. It is conceivable that this type of financial credits can be shared among co-workers in the same way as we axiomatically described above, and accordingly taken into account for academic ranking. A future possibility is to study how to obtain new quantitative features and rationally combine both quantitative and qualitative features in the web-mining and axiomatic framework.

In conclusion, we have integrated the axiomatic approach and the web technology to analyze the largest amount of scientific publications in the field of computer science for departmental ranking. The proposed axiomatic indices and self-citation exclusion scheme have corrected the subjective bias of the current ranking systems. Our data is clean and authoritative. Our work suggests a new concept of academic ranking that is content-wise rich, mathematically rigorous, and dynamically accessible.